\documentclass[aps,prd,reprint,footnoteinbib,longbibliography]{revtex4-2}
\pdfoutput=1
\usepackage[T1]{fontenc}
\usepackage{bm,amsmath,amssymb,color,bbm,mathrsfs} 
\usepackage[pdftex]{graphicx}
\usepackage[colorlinks=true, pdfstartview=FitH, linkcolor=blue, citecolor=blue, urlcolor=blue]{hyperref}
\usepackage{slashed}

\DeclareMathOperator\diag{diag}
\DeclareMathOperator\Tr{Tr}

\DeclareMathOperator\rme{\mathrm{e}}

\newcommand{\der}{\partial}
\renewcommand{\bar}[1]{\overline{#1}}

\newcommand{\bep}{\begin{pmatrix}} 
\newcommand{\eep}{\end{pmatrix}}

\newcommand{\SU}{\text{SU}}
\newcommand{\SO}{\text{SO}}
\renewcommand{\O}{\text{O}}
\newcommand{\U}{\text{U}}
\newcommand{\1}{\mathbbm{1}}
\newcommand{\RR}{\mathbb{R}}

\renewcommand{\epsilon}{\varepsilon}
\newcommand{\rmd}{\mathrm{d}}
\newcommand{\A}{\mathbb{A}}
\newcommand{\B}{\mathbb{B}}
\newcommand{\M}{\hat{M}}

\def\ba#1\ea{\begin{align}#1\end{align}}
\def\mkakko#1{\left( #1 \right)}

\def\kkakko#1{\left[ #1 \right]}

\begin{document}
\title{Chiral random matrix theory for colorful quark-antiquark condensates}

\author{Takuya Kanazawa}
\affiliation{Research and Development Group, Hitachi, Ltd., Kokubunji, Tokyo 185-8601, Japan}
\allowdisplaybreaks

\begin{abstract}
In QCD at high density, the color-octet quark-antiquark condensate $\langle\bar\psi\gamma_0(\lambda^A)_C (\lambda^A)_F\psi\rangle$ is generally nonzero and dynamically breaks the $\SU(3)_C\times \SU(3)_L\times\SU(3)_R$ symmetry down to the diagonal $\SU(3)_V$. We evaluate this condensate in the mean-field approximation and find that it is of order $\mu\Delta^2\log(\mu/\Delta)$ where $\Delta$ is the BCS gap of quarks. Next we propose a novel non-Hermitian chiral random matrix theory that describes the formation of colorful quark-antiquark condensates. We take the microscopic large-$N$ limit and find that three phases appear depending on the parameter of the model. They are the color-flavor locked phase, the polar phase, and the normal phase. We rigorously derive the effective theory of Nambu-Goldstone modes and determine the quark-mass dependence of the partition function. 
\end{abstract}
\maketitle

\section{Introduction}

Understanding confinement and chiral symmetry breaking in the QCD vacuum is a grand challenge in nuclear and hadron physics. It has been established that the QCD vacuum hosts a nonvanishing chiral condensate $\langle\bar\psi\psi\rangle$ which breaks the $\SU(N_f)_L\times\SU(N_f)_R$ chiral symmetry down to $\SU(N_f)_V$. In principle, one can also envision other condensates that lead to the same pattern of symmetry breaking. For example, the quark-gluon mixed condensate $\langle\bar\psi\sigma_{\mu\nu}G_{\mu\nu}\psi\rangle$ has been measured on the lattice \cite{Kremer:1987ve,Doi:2002wk}. It plays an important role in the QCD sum rule. In 1998, Stern pointed out a theoretical possibility that $\langle\bar\psi\psi\rangle=0$ and chiral symmetry is instead broken by a four-quark condensate \cite{Stern:1997ri,Stern:1998dy} (see also \cite{Harada:2009nq,Kanazawa:2015kca,Kanazawa:2016nlh,Yamaguchi:2018xse}). This interesting scenario was later ruled out by exact QCD inequalities \cite{Kogan:1998zc} at zero chemical potential and by the 't~Hooft anomaly matching condition \cite{Tanizaki:2018wtg} at nonzero chemical potential and zero temperature. In 1999, Wetterich proposed that a color-octet quark-antiquark condensate $\langle\bar\psi (\lambda^A)_C (\lambda^B)_F^T \psi\rangle\propto \delta_{AB}$ can provide a remarkably simple description of nonperturbative features of the QCD vacuum \cite{Wetterich:1999vd,Wetterich:1999sh,Wetterich:2000pp,Wetterich:2000ky} (see also \cite{Berges:2000nb,Berges:2000dh,Schafer:2001hr,Alford:2002rz,Gies:2006nz}). Here $(\lambda^A)_C$ denote the generators of color $\SU(3)$ and $(\lambda^B)_F$ the generators of flavor $\SU(3)$. Such a condensate locks $\SU(3)_L\times\SU(3)_R\times\SU(3)_C$ to the diagonal $\SU(3)_V$ subgroup, and consequently, quarks and gluons acquire nonzero masses. Their quantum numbers match those of baryons and vector mesons. Astoundingly, all physical particles in this phase carry integer electric charges. This beautiful Higgs description of confinement in QCD is in line with the well-known complementarity between a confining phase and a Higgs phase \cite{Fradkin:1978dv,Banks:1979fi}. It should be noted that Wetterich's phase has much in common with the color-flavor-locked (CFL) phase of three-flavor QCD at high density \cite{Alford:1998mk}, where colors and flavors are locked by diquark condensates. While the microscopic origin of the diquark condensates at high density is very clear, the physical mechanism that may give rise to the color-octet quark-antiquark condensate in the QCD vacuum is not well understood; both a one-gluon exchange interaction and an instanton-induced interaction are repulsive in this channel \cite{Alford:2002rz}. Nevertheless, in the CFL phase, the condensate $\langle\bar\psi (\lambda^A)_C (\lambda^A)_F^T \psi\rangle$ is expected to form, since it breaks no new symmetries \cite{Wetterich:2000pp,Berges:2000nb,Schafer:2001hr}. In \cite{Alford:2002rz} Alford~\textit{et~al.}~have performed a comprehensive analysis of the quark-antiquark pairing strength in various rotationally symmetric channels and found that the one-gluon exchange interaction is attractive in the channel $\bar\psi\gamma_0(\lambda^A)_C (\lambda^B)_F\psi$ and its pseudo-scalar partner. For the same reason as above, we expect that the color-flavor-locking condensate $\langle\bar\psi\gamma_0(\lambda^A)_C (\lambda^A)_F\psi\rangle$ would generally assume a nonzero value in the CFL phase. However, to the best of our knowledge, an explicit calculation of this condensate has not been performed to date.

It is widely accepted that statistical properties of the Dirac eigenvalues in a chirally broken phase of QCD are governed by chiral random matrix theory (RMT) \cite{Shuryak:1992pi,Verbaarschot:1993pm,Verbaarschot:2000dy,Verbaarschot:2005rj}. The precise  agreement between predictions of RMT and the numerically computed Dirac eigenvalues on the lattice is normally considered as a smoking gun for chiral symmetry breaking \cite{BerbenniBitsch:1997tx}. The connection between QCD and RMT holds at zero and small chemical potential \cite{Osborn:2004rf,Verbaarschot:2005rj,Akemann:2007rf}. In QCD-like theories such as two-color QCD, this connection can be extended to an arbitrarily large chemical potential \cite{Kanazawa:2009en,Akemann:2010tv,Kanazawa:2011tt,Kanazawabook,Kanazawa:2014lga}. There are also attempts to apply chiral RMT to color-superconducting phases with nonzero diquark condensates \cite{Vanderheyden:1999xp,Vanderheyden:2000ti,Pepin:2000pv,Sano:2011xs,Kanazawa:2020dju,Kanazawa:2020ktn}. We note in passing that flavor symmetry breaking in three-dimensional QCD can also be described by RMT with no chiral structure \cite{Verbaarschot:1994ip,Damgaard:1997pw,Kanazawa:2019oxu}. The interpolation between chiral RMT and non-chiral RMT has been studied extensively \cite{Damgaard:2010cz,Akemann:2010em,Akemann:2011kj,Kieburg:2011uf,Kanazawa:2018okt,Kanazawa:2018kbo,Kanazawa:2018ufq}.  

Inspired by Wetterich's work \cite{Wetterich:1999vd,Wetterich:1999sh,Wetterich:2000pp,Wetterich:2000ky}, in this paper we propose a new chiral RMT that describes color superconductivity due to the onset of the adjoint quark-antiquark condensate $\langle\bar\psi\gamma_0(\lambda^A)_C (\lambda^B)_F\psi\rangle$. As bold idealization, we shall ignore the chiral condensate and the diquark condensate that are predominant at low and high density, respectively. In this regard we admit that the proposed RMT is not of direct phenomenological relevance for the phase diagram of QCD \footnote{It is tempting to ask whether QCD at finite density can accommodate a hypothetical phase where colors and flavors are locked through the condensate $\protect\langle\bar\psi\gamma_0(\lambda^A)_C (\lambda^A)_F\psi\protect\rangle\ne0$ while $\protect\langle\bar\psi\psi\protect\rangle=\protect\langle\psi\psi\protect\rangle=0$. In this scenario, the anomaly-free axial symmetry $(\mathbb{Z}_{2N_f})_{\rm A}$ remains unbroken. This kind of symmetry breaking is strictly forbidden in QCD \cite{Kogan:1998zc,Tanizaki:2018wtg}.}. 
However we think it is a fruitful endeavor to widen the potential applicability of chiral RMT by searching for novel symmetry breaking patterns that have not been reported in the literature of RMT yet. 

This paper is structured as follows. In section~\ref{sc:qqcfl} we evaluate the color-flavor-locking quark-antiquark condensate in the CFL phase in the mean-field approximation, and show that it is of order $\mu\Delta^2$ and grows monotonically with $\mu$, in contrast to the ordinary chiral condensate which is highly suppressed at large $\mu$ \cite{Schafer:2002ty}. In section~\ref{sc:mm} we introduce a new matrix model and perform a Hubbard-Stratonovich transformation. In section~\ref{sc:22} we focus on the case of two colors and two flavors. We take the microscopic large-$N$ limit with $N$ the matrix size and, by varying a parameter of the matrix model, find three distinct phases: the normal phase, the polar phase and the color-flavor locked phase, which we refer to as \emph{the adjoint CFL phase} to distinguish it from the ordinary CFL phase with diquark condensates. We rigorously derive the large-$N$ effective theory for the Nambu-Goldstone modes in the polar phase and the adjoint CFL phase, and determine the quark-mass dependence of the partition function. In section~\ref{sc:conc} we end with a summary and outlook.

\section{\label{sc:qqcfl}Quark-antiquark condensate in the CFL phase}

The purpose of this section is to evaluate the magnitude of the condensate
\ba
	\langle\bar\psi\gamma_0(\lambda^A)_C (\lambda^A)_F\psi \rangle
	\label{eq:cond}
\ea
in the CFL phase of QCD with three colors and three flavors in the chiral limit. In the following, we label colors by $a,b,\cdots\in\{1,2,3\}$ and flavors by $f,g,\cdots\in\{1,2,3\}$. The indices $\A,\B,\cdots$ run from 0 to 8 and $A,B,\cdots$ from 1 to 8. The Gell-Mann matrices are normalized as $\Tr(\lambda^\A\lambda^\B)=2\delta_{\A\B}$ where $\lambda^0 \equiv \sqrt{\frac{2}{3}}\1_3$. 

The mean-field Lagrangian for the CFL phase is given, in the Euclidean setup \footnote{The gamma matrices are Hermitian and satisfy $\{\gamma_\mu,\gamma_\nu\}=2\delta_{\mu\nu}\1_4$.}, by
\ba
	\mathcal{L} & = \bar\psi_{af}(\slashed{\der}-\mu\gamma_0)\psi_{af}
	+ \frac{1}{2}\psi_{af}^T C\gamma_5 \psi_{bg} \epsilon_{abI}\epsilon_{fgI}\Delta + \text{h.c.}
	\label{eq:Lmf}
\ea
To simplify the calculation we switch to the CFL basis
\ba
	\psi_{af}=\frac{1}{\sqrt{2}}\lambda^\A_{af}\psi^\A\,.
\ea
Then
\ba
	\mathcal{L} & = {\bar{\psi^\A}}(\slashed{\der}-s_{\A}\mu\gamma_0)\psi^\A
	+ \frac{1}{2}(\psi^{\A})^TC\gamma_5\psi^\A \Delta_\A + \text{h.c.},
\ea
where 
\ba
	\Delta_\A & = \begin{cases}2\Delta\qquad \text{for}\quad \A=0\\-\Delta\qquad \!\!\text{for}\quad \A=1,\cdots,8\end{cases}
\ea
and
\ba
	s_\A & = \frac{1}{2}\Tr[\lambda^\A(\lambda^\A)^T]
	\\
	& = (1,1,-1,1,1,-1,1,-1,1)\,.
\ea
Then
\ba
	\bar\psi\gamma_0(\lambda^A)_C (\lambda^A)_F\psi 
	& = 2 \bar\psi_{af}\gamma_0\psi_{fa} - \frac{2}{3}\bar\psi_{af}\gamma_0 \psi_{af}
	\\
	& = 2\bar{\psi^\A}\gamma_0\psi^\A - \frac{2}{3} s_\A \bar{\psi^\A}\gamma_0\psi^\A
	\\
	& = \frac{4}{3}(\bar{\psi^0}\gamma_0\psi^0-\bar{\psi^1}\gamma_0\psi^1)\,,
\ea
where we have used the fact that $\bar{\psi^A}\gamma_0\psi^A = s_A \bar{\psi^1}\gamma_0\psi^1$ (no summation over $A$ on the LHS). The number density $\bar{\psi^1}\gamma_0\psi^1$ can be evaluated either by using the propagator
\ba
	\langle\psi^1\bar{\psi^1}\rangle & = -(i\slashed{p}+\mu\gamma_0)
	\frac{1}{\Delta^2-(i\slashed{p}-\mu\gamma_0)(i\slashed{p}+\mu\gamma_0)}
\ea
or by taking the derivative of the logarithm of the functional determinant by $\mu$. The result reads
\ba
	\langle\bar{\psi^1}\gamma_0\psi^1\rangle 
	= \frac{\der}{\der \mu}\int\frac{\rmd^3 p}{(2\pi)^3}[E_{+}(\mathbf{p})+E_{-}(\mathbf{p})]
	\label{eq:23423432}
\ea
where 
\ba
	E_\pm(\mathbf{p}) & \equiv \sqrt{\Delta^2+(|\mathbf{p}|\pm \mu)^2}\,.
\ea
The momentum integral is UV divergent and we impose a cutoff $\Lambda$. We obtain (assuming $\mu>0$)
\ba
	\langle\bar{\psi^1}\gamma_0\psi^1\rangle & = \frac{\mu^3}{2\pi^2}
	f\mkakko{\frac{\Delta}{\mu},\frac{\Lambda}{\mu}},
\ea
with
\ba
	f(a,b) & \equiv \int_0^{b}\rmd x\;x^2\mkakko{
		\frac{1-x}{\sqrt{(x-1)^2+a^2}} + \frac{1+x}{\sqrt{(x+1)^2+a^2}}
	}
	\\
	& \approx \begin{cases}
	2a^2\log b \;\quad \qquad \text{for}~~b\gg 1
	\\
	\displaystyle \frac{2}{3} - 2a^2\log a \quad \,\text{for}~~ a\ll 1 ~~\text{and}~~ b>1
	\end{cases}
	\hspace{-10pt}
\ea
Therefore
\ba
	& \langle \bar\psi\gamma_0(\lambda^A)_C (\lambda^A)_F\psi \rangle 
	\notag
	\\
	=\; & \frac{2\mu^3}{3\pi^2}\kkakko{
		f\mkakko{\frac{2\Delta}{\mu},\frac{\Lambda}{\mu}} 
		- f\mkakko{\frac{\Delta}{\mu},\frac{\Lambda}{\mu}}
	}
	\\
	\approx \; & \begin{cases}
	\displaystyle 
	\frac{4}{\pi^2}\mu\Delta^2\log\frac{\Lambda}{\mu} 
	\qquad \text{for}~~\Lambda \gg \mu
	\\
	\displaystyle 
	\frac{4}{\pi^2}\mu\Delta^2\log\frac{\mu}{\Delta} \qquad \!\text{for}~~\Delta\ll\mu~~\text{and}~~\Lambda>\mu
	\end{cases}
\ea
This is the main result of this section. As expected on symmetry grounds, it does not vanish in the CFL phase. It grows with $\mu$ monotonically, in contrast to the chiral condensate $\langle\bar\psi\psi\rangle$ which vanishes identically in the mean-field approximation. (It receives contributions from instantons \cite{Schafer:2002ty}.) 

By replacing $\der/\der \mu$ with $\der/\der \Delta$ in \eqref{eq:23423432}, the diquark condensate $\langle\psi\psi\rangle$ is obtained as $\sim \mu^2\Delta\log(\mu/\Delta)$ for $\Delta\ll\mu$ \footnote{The fact that the octet condensate is roughly $\mu\Delta^2$ while the diquark condensate is $\mu^2\Delta$ has an intuitive explanation. The octet condensate has the quantum number of charge density so it should change sign when we flip $\mu\to -\mu$. Note also that, as is clear from \eqref{eq:Lmf}, if we view $\Delta$ as a spurious external field, it is charged under $\U(1)_B$. The octet condensate is singlet under $\U(1)_B$, so it should be proportional to $|\Delta|^2$. We thus expect that the octet condensate is roughly $\mu|\Delta|^2$. On the other hand, the diquark condensate is charged under $\U(1)_B$, so it should be proportional to $\Delta^*$. Next, note that the chemical potential $\mu\gamma_0$ can be expressed as a vector field $A_\nu\gamma_\nu$ with $A_\nu=(\mu,0,0,0)$. Since the diquark condensate is a Lorentz scalar, it must depend on $A_\nu$ only through $A^2$. Thus we expect that the diquark condensate is roughly $\mu^2 \Delta^*$.}, so we get the hierarchy of scales
\ba
	\frac{\langle \bar\psi\gamma_0(\lambda^A)_C (\lambda^A)_F\psi \rangle }{|\langle\psi\psi\rangle|} \propto \frac{\Delta}{\mu} \ll 1\,.
\ea

\section{\label{sc:mm}The matrix model}

Next we proceed to the random matrix analysis. In the following we assume that the number of colors and flavors are equal, i.e.,
\ba
	N_c = N_f = n\,.
\ea
The indices $\A,\B,\cdots$ run from $0$ to $n^2-1$ and $A,B,\cdots$ from $1$ to $n^2-1$. $\{T^\A\}$ are the generators of $\U(n)$ in the fundamental representation, normalized as $\Tr(T^\A T^\B)=2\delta_{\A\B}$. 

The new RMT we propose in this paper is defined by the partition function
\begin{widetext}
\ba
	Z(m_f,v) & = \int_{-\infty}^{\infty} \rmd x \int_{-\infty}^{\infty} \rmd y 
	\int \rmd A \int \rmd B \int \rmd V \int \rmd W 
	\int \prod_{\A=0}^{n^2-1}\rmd X^{\A}\int \prod_{\A=0}^{n^2-1}\rmd Y^{\A}
	\notag
	\\
	& \quad \times \exp\kkakko{
		-\frac{Nn^2}{4}(x^2+y^2)-\frac{N}{4}\Tr(A^2+B^2)-\frac{Nn}{4}\Tr(V^2+W^2)
		-\frac{Nn}{2}\Tr[(X^\A)^2+(Y^\A)^2]
	}
	\notag
	\\
	& \quad \times \prod_{f=1}^{n}\det\kkakko{\mathscr{D}+
	\begin{pmatrix}m_f\1_{Nn} & v\1_{Nn}\\ v\1_{Nn} & m^*_f\1_{Nn}\end{pmatrix}} \,,
	\label{eq:defZ}
\ea
\end{widetext}
where $A,B,X^\A,Y^\A$ are $N\times N$ Hermitian matrices, and $V$ and $W$ are $n\times n$ Hermitian matrices. The ``Dirac operator'' $\mathscr{D}$ is a non-Hermitian $2Nn\times 2Nn$ matrix defined as 
\begin{gather}
	\mathscr{D} = \begin{pmatrix}0&D_R\\D_L&0\end{pmatrix}
	\label{eq:defDD}
	\\
	D_L \equiv x \1_N\otimes \1_n + iA\otimes\1_n + i \1_N\otimes V + X^\A \otimes T^\A
	\hspace{-7pt}
	\\
	D_R \equiv y \1_N\otimes \1_n + iB\otimes\1_n + i \1_N\otimes W + Y^\A \otimes T^\A
	\hspace{-7pt}
\end{gather}
In \eqref{eq:defZ}, $\{m_f\}$ are quark masses that break chiral symmetry, and $v$ is a real parameter that conserves chiral symmetry. The importance of this mysterious parameter will become clear later. In the chiral limit the model possesses a symmetry
\ba
	& [\U(1)\times\SU(n)_C\times\SU(n)_F]_L 
	\notag
	\\
	\times & [\U(1)\times\SU(n)_C\times\SU(n)_F]_R \,.
\ea
The left color $[\SU(n)_C]_L$ and the right color $[\SU(n)_C]_R$ are locked to $[\SU(n)_C]_{L+R}$ by nonzero quark masses. 

Let us introduce quarks $\psi^\alpha_{Laf},\psi^\alpha_{Raf}$ and antiquarks $\bar\psi^\alpha_{Laf},\bar\psi^\alpha_{Raf}$, where $a\in\{1,\cdots,n\}$ is color, $f\in\{1,\cdots,n\}$ is flavor, and $\alpha\in\{1,\cdots,N\}$. Then, with the $n\times n$ mass matrix $M\equiv \diag(m_f)$ we have
\begin{widetext}
\ba
	Z & = \int \rmd \bar\psi_R \rmd \bar\psi_L \rmd \psi_R \rmd \psi_L 
	\int_{-\infty}^{\infty} \rmd x \int_{-\infty}^{\infty} \rmd y 
	\int \rmd A \int \rmd B \int \rmd V \int \rmd W 
	\int \prod_{\A=0}^{n^2-1}\rmd X^{\A}\int \prod_{\A=0}^{n^2-1}\rmd Y^{\A}
	\notag
	\\
	& \quad \times \exp\kkakko{
		-\frac{Nn^2}{4}(x^2+y^2)-\frac{N}{4}\Tr(A^2+B^2)-\frac{Nn}{4}\Tr(V^2+W^2)
		-\frac{Nn}{2}\Tr[(X^\A)^2+(Y^\A)^2]
	}
	\notag
	\\
	& \quad \times \exp\kkakko{
		\begin{pmatrix}\bar\psi_R \\ \bar\psi_L \end{pmatrix}^\alpha_{af}
		\begin{pmatrix}
		m_f \delta_{\alpha\beta}\delta_{ab} & (y+v)\delta_{\alpha\beta}\delta_{ab}+iB_{\alpha\beta}\delta_{ab}+i\delta_{\alpha\beta}W_{ab}+Y^\A_{\alpha\beta}T^\A_{ab}
		\\
		(x+v)\delta_{\alpha\beta}\delta_{ab} + iA_{\alpha\beta}\delta_{ab}+i\delta_{\alpha\beta}V_{ab}+X^\A_{\alpha\beta}T^\A_{ab} & m_f^* \delta_{\alpha\beta}\delta_{ab}
		\end{pmatrix}
		\begin{pmatrix}\psi_L \\ \psi_R\end{pmatrix}^\beta_{bf}
	}
	\\
	& = \int \rmd \bar\psi_R \rmd \bar\psi_L \rmd \psi_R \rmd \psi_L 
	\int_{-\infty}^{\infty} \rmd x \int_{-\infty}^{\infty} \rmd y 
	\int \rmd A \int \rmd B \int \rmd V \int \rmd W 
	\int \prod_{\A=0}^{n^2-1}\rmd X^{\A}\int \prod_{\A=0}^{n^2-1}\rmd Y^{\A}
	\notag
	\\
	& \quad \times \exp\kkakko{
		-\frac{Nn^2}{4}(x^2+y^2)-\frac{N}{4}\Tr(A^2+B^2)-\frac{Nn}{4}\Tr(V^2+W^2)
		-\frac{Nn}{2}\Tr[(X^\A)^2+(Y^\A)^2]
	}
	\notag
	\\
	& \quad \times \exp\Big[
		(x+v) \bar\psi^\alpha_{Laf}\psi^\alpha_{Laf} + i \bar\psi^\alpha_{Laf}A_{\alpha\beta}\psi^\beta_{Laf}
		+ i \bar\psi^\alpha_{Laf}V_{ab}\psi^\alpha_{Lbf}
		+ \bar\psi^\alpha_{Laf}X^\A_{\alpha\beta}T^\A_{ab}\psi^\beta_{Lbf}
		+ (y+v) \bar\psi^\alpha_{Raf}\psi^\alpha_{Raf} + i \bar\psi^\alpha_{Raf}B_{\alpha\beta}\psi^\beta_{Raf}
	\notag
	\\
	& \quad 
		+ i \bar\psi^\alpha_{Raf}W_{ab}\psi^\alpha_{Rbf}
		+ \bar\psi^\alpha_{Raf}Y^\A_{\alpha\beta}T^\A_{ab}\psi^\beta_{Rbf}
		+ \bar\psi_{Raf}^\alpha M_{fg} \psi^\alpha_{Lag} + \bar\psi^\alpha_{Laf}(M^\dagger)_{fg}\psi^\alpha_{Rag}
	\Big] \,.
\ea
Now it is straightforward albeit tedious to integrate out all the Gaussian variables, which yields
\ba
	Z & \propto \int \rmd \bar\psi_R \rmd \bar\psi_L \rmd \psi_R \rmd \psi_L 
	\exp\bigg[
		\bigg\{
		\frac{1}{Nn^2}(\bar\psi^\alpha_{Laf}\psi^\alpha_{Laf})^2 
		- \frac{1}{N}\bar\psi^\beta_{Laf}\psi^\alpha_{Laf}\bar\psi^\alpha_{Lbg}\psi^\beta_{Lbg}
		-\frac{1}{Nn}\bar\psi^\alpha_{Lbf}\psi^\alpha_{Laf}\bar\psi^\beta_{Lag}\psi^\beta_{Lbg}
	\notag
	\\
	& \quad 
		+ \frac{1}{2Nn}\bar\psi^\beta_{Laf}T^\A_{ab}\psi^\alpha_{Lbf}
		\bar\psi^\alpha_{Lcg}T^\A_{cd}\psi^\beta_{Ldg}
		+ v \bar\psi^\alpha_{Laf}\psi^\alpha_{Laf}
		\bigg\}
		+ (L\leftrightarrow R) 
		+ \bar\psi_{Raf}^\alpha M_{fg} \psi^\alpha_{Lag} + \bar\psi^\alpha_{Laf}(M^\dagger)_{fg}\psi^\alpha_{Rag}
	\bigg]
	\\
	& = \int \rmd \bar\psi_R \rmd \bar\psi_L \rmd \psi_R \rmd \psi_L 
	\exp\bigg[
		\bigg\{
		\frac{1}{Nn^2}(\bar\psi^\alpha_{Laf}\psi^\alpha_{Laf})^2 
		- \frac{1}{N}\bar\psi^\beta_{Laf}\psi^\alpha_{Laf}\bar\psi^\alpha_{Lbg}\psi^\beta_{Lbg}
		-\frac{1}{Nn}\bar\psi^\alpha_{Lbf}\psi^\alpha_{Laf}\bar\psi^\beta_{Lag}\psi^\beta_{Lbg}
	\notag
	\\
	& \quad 
		+ \frac{1}{Nn}\bar\psi^\beta_{Laf}\psi^\alpha_{Lbf}
		\bar\psi^\alpha_{Lbg}\psi^\beta_{Lag}
		+ v \bar\psi^\alpha_{Laf}\psi^\alpha_{Laf}
		\bigg\}
		+ (L\leftrightarrow R) 
		+ \bar\psi_{Raf}^\alpha M_{fg} \psi^\alpha_{Lag} + \bar\psi^\alpha_{Laf}(M^\dagger)_{fg}\psi^\alpha_{Rag}
	\bigg] \,,
\ea
where we have used $T^\A_{ab} T^\A_{cd}=2\delta_{ad}\delta_{bc}$. Rearranging terms, we have
\ba
	Z & \propto 
	\int \rmd \bar\psi_R \rmd \bar\psi_L \rmd \psi_R \rmd \psi_L 
	\exp\bigg[ \frac{1}{N}\bigg(
		\bar\psi^\alpha_{Laf}\psi^\alpha_{Lbg}\bar\psi^\beta_{Lbg}\psi^\beta_{Laf}
		-\frac{1}{n}\bar\psi^\alpha_{Laf}\psi^\alpha_{Lbf}\bar\psi^\beta_{Lbg}\psi^\beta_{Lag}
		- \frac{1}{n}\bar\psi^\alpha_{Laf}\psi^\alpha_{Lag}\bar\psi^\beta_{Lbg}\psi^\beta_{Lbf}
	\notag
	\\
	& \quad 
		+\frac{1}{n^2}\bar\psi^\alpha_{Laf}\psi^\alpha_{Laf}\bar\psi^\beta_{Lbg}\psi^\beta_{Lbg}
		\bigg) 
		+ v \bar\psi^\alpha_{Laf}\psi^\alpha_{Laf}
		+ (L\leftrightarrow R) 
		+ \bar\psi_{Raf}^\alpha M_{fg} \psi^\alpha_{Lag} + \bar\psi^\alpha_{Laf}(M^\dagger)_{fg}\psi^\alpha_{Rag}
	\bigg]
	\\
	& = \int \rmd \bar\psi_R \rmd \bar\psi_L \rmd \psi_R \rmd \psi_L 
	\exp\bigg[ \frac{1}{N}
	\bar\psi^\alpha_{Laf}\psi^\alpha_{Lbg}\bar\psi^\beta_{La'f'}\psi^\beta_{Lb'g'}
	\bigg(
		\delta_{ab'}\delta_{a'b}\delta_{fg'}\delta_{f'g}
		-\frac{1}{n}\delta_{ab'}\delta_{a'b}\delta_{fg}\delta_{f'g'}
		-\frac{1}{n}\delta_{ab}\delta_{a'b'}\delta_{fg'}\delta_{f'g}
	\notag
	\\
	& \quad 
		+\frac{1}{n^2}\delta_{ab}\delta_{a'b'}\delta_{fg}\delta_{f'g'}
		\bigg)
		+ v \bar\psi^\alpha_{Laf}\psi^\alpha_{Laf}
		+ (L\leftrightarrow R) 
		+ \bar\psi_{Raf}^\alpha M_{fg} \psi^\alpha_{Lag} + \bar\psi^\alpha_{Laf}(M^\dagger)_{fg}\psi^\alpha_{Rag}
	\bigg]
	\\
	& = \int \rmd \bar\psi_R \rmd \bar\psi_L \rmd \psi_R \rmd \psi_L 
	\exp\bigg[ \frac{1}{N}
	\bar\psi^\alpha_{Laf}\psi^\alpha_{Lbg}\bar\psi^\beta_{La'f'}\psi^\beta_{Lb'g'}
	\bigg(
		\delta_{ab'}\delta_{a'b}-\frac{1}{n}\delta_{ab}\delta_{a'b'}
	\bigg)
	\bigg(
		\delta_{fg'}\delta_{f'g}-\frac{1}{n}\delta_{fg}\delta_{f'g'}
	\bigg)
	\notag
	\\
	& \quad 
		+ v \bar\psi^\alpha_{Laf}\psi^\alpha_{Laf}
		+ (L\leftrightarrow R) 
		+ \bar\psi_{Raf}^\alpha M_{fg} \psi^\alpha_{Lag} + \bar\psi^\alpha_{Laf}(M^\dagger)_{fg}\psi^\alpha_{Rag}
	\bigg]
	\\
	& = \int \rmd \bar\psi_R \rmd \bar\psi_L \rmd \psi_R \rmd \psi_L 
	\exp\bigg[ \frac{1}{4N}
	\bar\psi^\alpha_{Laf}\psi^\alpha_{Lbg}\bar\psi^\beta_{La'f'}\psi^\beta_{Lb'g'}
	(T^A_C)_{ab}(T^A_C)_{a'b'}(T^B_F)_{fg}(T^B_F)_{f'g'} 
	+ v \bar\psi^\alpha_{Laf}\psi^\alpha_{Laf}
	\notag
	\\
	& \quad 
		+ (L\leftrightarrow R) 
		+ \bar\psi_{Raf}^\alpha M_{fg} \psi^\alpha_{Lag} + \bar\psi^\alpha_{Laf}(M^\dagger)_{fg}\psi^\alpha_{Rag}
	\bigg]
	\\
	& = \int \rmd \bar\psi_R \rmd \bar\psi_L \rmd \psi_R \rmd \psi_L 
	\exp\bigg[ \frac{1}{4N}
	\Big(
		\bar\psi^\alpha_{Laf}(T^A_C)_{ab}(T^B_F)_{fg}\psi^\alpha_{Lbg}
	\Big)\Big( 
		\bar\psi^\beta_{La'f'}(T^A_C)_{a'b'}(T^B_F)_{f'g'}\psi^\beta_{Lb'g'}
	\Big) + v \bar\psi^\alpha_{Laf}\psi^\alpha_{Laf}
	\notag
	\\
	& \quad 
		+ (L\leftrightarrow R) 
		+ \bar\psi_{Raf}^\alpha M_{fg} \psi^\alpha_{Lag} + \bar\psi^\alpha_{Laf}(M^\dagger)_{fg}\psi^\alpha_{Rag}
	\bigg]
\ea
where $T_C^A$ and $T_F^A$ are the generators of color $\SU(n)$ and flavor $\SU(n)$, respectively. 

To bilinearize the quartic interaction we insert the constant factor
\ba
	\int \rmd \Omega^L \exp\bigg[-N\mkakko{\Omega^L_{AB}-\frac{1}{2N}\bar\psi^\alpha_{Laf}(T^A_C)_{ab}(T^B_F)_{fg}\psi^\alpha_{Lbg}}^2\bigg]
	\\
	\times 
	\int \rmd \Omega^R \exp\bigg[-N\mkakko{\Omega^R_{AB}-\frac{1}{2N}\bar\psi^\alpha_{Raf}(T^A_C)_{ab}(T^B_F)_{fg}\psi^\alpha_{Rbg}}^2\bigg]
\ea
where $\Omega^L$ and $\Omega^R$ are $(n^2-1)\times(n^2-1)$ real matrices with no symmetry constraint. This yields 
\ba
	Z & \propto \int \rmd \Omega^L \int \rmd \Omega^R \exp\mkakko{-N\Tr[(\Omega^L)^T\Omega^L+(\Omega^R)^T\Omega^R]}\int \rmd \bar\psi_R \rmd \bar\psi_L \rmd \psi_R \rmd \psi_L 
	\notag
	\\
	& \quad \times \exp\kkakko{
		\bar\psi^\alpha_{Laf}\Omega^L_{AB}(T^A_C)_{ab}(T^B_F)_{fg}\psi^\alpha_{Lbg}
		+ v \bar\psi^\alpha_{Laf}\psi^\alpha_{Laf}
		+ (L\leftrightarrow R) 
		+ \bar\psi_{Raf}^\alpha M_{fg} \psi^\alpha_{Lag} + \bar\psi^\alpha_{Laf}(M^\dagger)_{fg}\psi^\alpha_{Rag}
	}
	\\
	& = \int \rmd \Omega^L \int \rmd \Omega^R \exp\mkakko{-N\Tr[(\Omega^L)^T\Omega^L+(\Omega^R)^T\Omega^R]}\int \rmd \bar\psi_R \rmd \bar\psi_L \rmd \psi_R \rmd \psi_L 
	\notag
	\\
	& \quad \times \exp\kkakko{
		\begin{pmatrix}\bar\psi_R \\ \bar\psi_L\end{pmatrix}^\alpha_{af}
		\begin{pmatrix}
		\delta_{ab}M_{fg} & \Omega^R_{AB}(T^A_C)_{ab}(T^B_F)_{fg}+ v\delta_{ab}\delta_{fg}
		\\
		\Omega^L_{AB}(T^A_C)_{ab}(T^B_F)_{fg} + v\delta_{ab}\delta_{fg} & \delta_{ab}(M^\dagger)_{fg}
		\end{pmatrix}
		\begin{pmatrix}\psi_L\\\psi_R\end{pmatrix}^\alpha_{bg}
	}
	\\
	&= \int \rmd \Omega^L \int \rmd \Omega^R \exp\mkakko{-N\Tr[(\Omega^L)^T\Omega^L+(\Omega^R)^T\Omega^R]}
	{\det}^N \begin{pmatrix}
		\1_n\otimes M & \Omega^R_{AB}T^A_C \otimes T^B_F + v \1_n\otimes \1_n
		\\
		\Omega^L_{AB}T^A_C \otimes T^B_F + v \1_n\otimes \1_n & \1_n \otimes M^\dagger
		\end{pmatrix}. 
	\label{eq:924799}
\ea
\end{widetext}
This is an exact rewriting of the original partition function and so far no approximation has been made. To make the large-$N$ saddle-point analysis tractable, hereafter we set $n=2$.

\section{\label{sc:22}Two colors and two flavors}
\subsection{\boldmath Phase structure at large $N$}

\begin{figure}[tb]
	\centering
	\includegraphics[width=0.85\columnwidth]{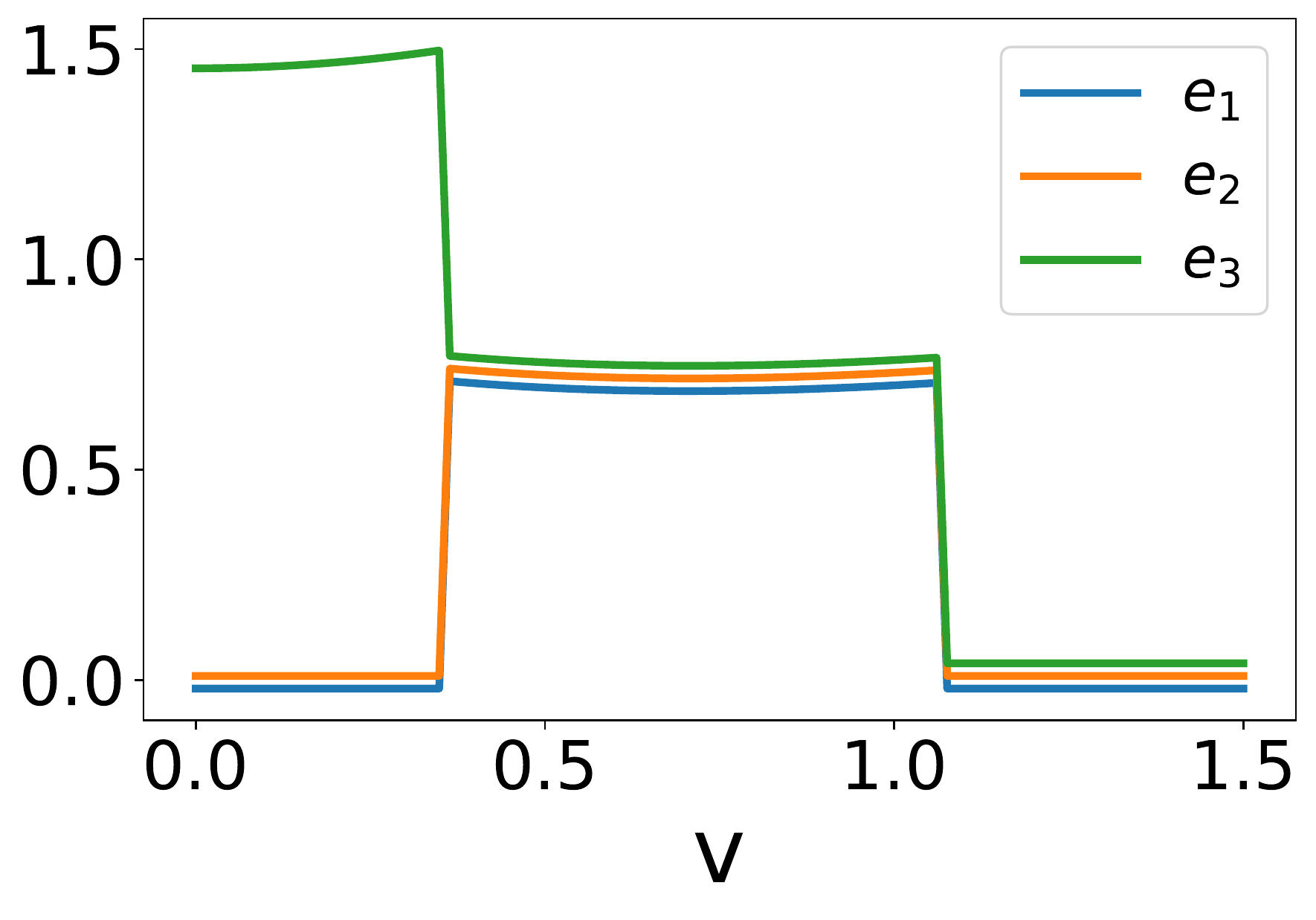}
	\caption{\label{fg:eee}The $v$-dependence of $\{e_k\}$ that maximize $F(v,\{e_k\})$. For better visibility, we slightly displaced the plots vertically so that they do not exactly coincide.}
\end{figure}

For $n=2$, $\Omega^{L,R}$ are $3\times 3$ matrices that transform as $(\mathbf{3},\mathbf{3})$ under $\SU(2)_C\times\SU(2)_F$ rotations, and $T^A_C$ and $T^A_F$ are just the Pauli matrices $\sigma^A(A=1,2,3)$. We are interested in the microscopic large-$N$ limit in which $M\sim 1/\sqrt{N}$ and $v\sim \O(1)$. From \eqref{eq:924799} we have for the partition function 
\ba
	Z & \propto \int \rmd \Omega^L \Big[
	\rme^{-\Tr[(\Omega^L)^T\Omega^L]}	
	\notag
	\\
	& \quad \times \det(\Omega^L_{AB}\sigma^A\otimes\sigma^B+v\1_2\otimes\1_2)
	\Big]^N
	\notag
	\\
	& \quad \times (L\leftrightarrow R)\,.
\ea
One can use the identity \footnote{We used SymPy, a Python library for symbolic mathematics \cite{10.7717/peerj-cs.103}.}
\ba
	& \det(\Omega_{AB}\sigma^A\otimes \sigma^B+v\1_2\otimes\1_2)
	\notag
	\\
	= \; & 2 \Tr[(\Omega^T \Omega)^2] - [\Tr(\Omega^T \Omega)]^2 - 8v \det \Omega 
	\notag
	\\
	& - 2v^2 \Tr(\Omega^T\Omega) + v^4\,.
\ea
By a rotation $\Omega\to O_1 \Omega O_2$ with $O_{1,2}\in\SO(3)$, $\Omega$ can be diagonalized as $\Omega=\diag(e_1,e_2,e_3)$. Our task is to find $\{e_k\}$ that maximizes the function
\ba
	F(v,\{e_k\}) & \equiv \rme^{-2(e_1^2+e_2^2+e_3^2)}
	\Big\{
		2(e_1^4+e_2^4+e_3^4) 
	\notag
	\\
	& \quad 
	- (e_1^2+e_2^2+e_3^2)^2 - 8v e_1e_2e_3 
	\notag
	\\
	& \quad 
	- 2v^2(e_1^2+e_2^2+e_3^2) + v^4 
	\Big\}^2.
\ea
Due to the trivial symmetry $F(v,\{e_k\})=F(-v,\{-e_k\})$ we may assume $v\geq 0$ without loss of generality. We numerically solved the maximization problem for varying $v$ and obtained $(e_1,e_2,e_3)$ as a function of $v$, as shown in Figure~\ref{fg:eee}. 
One can see two first-order transitions that separate three phases. In the first phase at small $v$, $e_1=e_2=0$ and $e_3>0$. This is an analogue of the polar phase of superfluid $^3$He \cite{HeliumBook}. In the second phase at intermediate $v$, we have $e_1=e_2=e_3$, i.e., $\Omega\propto \1_3$. This is a color-flavor locked phase, resembling the B phase of superfluid $^3$He \cite{HeliumBook}. Hereafter we call this phase \emph{the adjoint CFL phase}, to avoid confusion with the usual CFL phase. The final phase at large $v$ is a normal phase characterized by $e_1=e_2=e_3=0$. The phases are summarized below.
\vspace{5mm}\\
\begin{tabular}{|c|c|c|c|}
    \hline
    &$|v|<0.357$&$0.357<|v|<1.072$&$1.072<|v|$
    \\\hline \hline
    Phases & Polar & Adjoint CFL & Normal
    \\\hline 
    $\Omega$ & $\propto\diag(0,0,1)$ & $\propto\1_3$ & 0
    \\\hline 
    $\begin{array}{c}\text{Unbroken}\\\text{symmetry}\end{array}$ 
    &$\begin{array}{c}\SO(2)_C \\ \times \SO(2)_F\end{array}$& $\SU(2)_{C+F}$
    & $\begin{array}{c}\SU(2)_C \\ \times \SU(2)_F\end{array}$
    \\\hline
    \text{Coset space}
    & $S^2\times S^2$ & $\SO(3)$ & \text{None}
    \\\hline
\end{tabular}

\subsection{Effective theory of the polar phase}

Let us discuss the low-energy fluctuations in the polar phase. For simplicity we set $v=0$, for which the ground state is $\Omega^{L,R}=\diag(0,0,\sqrt{2})$. The soft fluctuations of color and flavor degrees of freedom can be parametrized, in terms of normalized three-component vectors $\overrightarrow{L_C},\overrightarrow{L_F},\overrightarrow{R_C},\overrightarrow{R_F}\in\RR^3$, as
\begin{gather}
	\Omega^L_{AB}(\sigma^A)_{ab}(\sigma^B)_{fg} = \sqrt{2}(\overrightarrow{L_C}\cdot \overrightarrow{\sigma})_{ab}(\overrightarrow{L_F}\cdot \overrightarrow{\sigma})_{fg}, 
	\\
	\Omega^R_{AB}(\sigma^A)_{ab}(\sigma^B)_{fg} = \sqrt{2}(\overrightarrow{R_C}\cdot \overrightarrow{\sigma})_{ab}(\overrightarrow{R_F}\cdot \overrightarrow{\sigma})_{fg}, 
	\\
	\overrightarrow{L_C}^2=\overrightarrow{L_F}^2=\overrightarrow{R_C}^2=\overrightarrow{R_F}^2=1.
\end{gather}
Thus, for $N\gg 1$,
\ba
	Z & \sim \int_{S^2} \rmd \overrightarrow{L_C}
	\int_{S^2} \rmd \overrightarrow{L_F}
	\int_{S^2} \rmd \overrightarrow{R_C}
	\int_{S^2} \rmd \overrightarrow{R_F}
	\notag
	\\
	& \times 
	{\det}^N \begin{pmatrix}
	(\overrightarrow{L_C}\cdot \overrightarrow{\sigma})\otimes(\overrightarrow{L_F}\cdot \overrightarrow{\sigma}) & \1_2\otimes M^\dagger/\sqrt{2}
		\\
		\1_2\otimes M/\sqrt{2} & (\overrightarrow{R_C}\cdot \overrightarrow{\sigma})\otimes(\overrightarrow{R_F}\cdot \overrightarrow{\sigma})
		\end{pmatrix}.
\ea
Using $(\overrightarrow{L_C}\cdot \overrightarrow{\sigma})^2=(\overrightarrow{L_F}\cdot \overrightarrow{\sigma})^2=(\overrightarrow{R_C}\cdot \overrightarrow{\sigma})^2=(\overrightarrow{R_F}\cdot \overrightarrow{\sigma})^2=\1_2$ we obtain
\begin{widetext}
\ba
	Z & \sim \int_{S^2} \rmd \overrightarrow{L_C}
	\int_{S^2} \rmd \overrightarrow{L_F}
	\int_{S^2} \rmd \overrightarrow{R_C}
	\int_{S^2} \rmd \overrightarrow{R_F}
	\notag
	\\
	& \quad \times 
	{\det}^N \begin{pmatrix}
		\1_4 & 
		[(\overrightarrow{L_C}\cdot \overrightarrow{\sigma})\otimes(\overrightarrow{L_F}\cdot \overrightarrow{\sigma})]
		\cdot [\1_2\otimes M^\dagger]/\sqrt{2}
		\\
		[(\overrightarrow{R_C}\cdot \overrightarrow{\sigma})\otimes(\overrightarrow{R_F}\cdot \overrightarrow{\sigma})] 
		\cdot [\1_2\otimes M]/\sqrt{2}
		& \1_4
	\end{pmatrix}
	\\
	& = \int_{S^2} \rmd \overrightarrow{L_C}
	\int_{S^2} \rmd \overrightarrow{L_F}
	\int_{S^2} \rmd \overrightarrow{R_C}
	\int_{S^2} \rmd \overrightarrow{R_F}~~{\det}^N\mkakko{
		\1_4 - \frac{1}{2}
		\Big\{
		(\overrightarrow{R_C}\cdot \overrightarrow{\sigma})\otimes[(\overrightarrow{R_F}\cdot \overrightarrow{\sigma})M]
		\Big\}\Big\{
		(\overrightarrow{L_C}\cdot \overrightarrow{\sigma}) \otimes[(\overrightarrow{L_F}\cdot \overrightarrow{\sigma})M^\dagger]\Big\}
	}.
\ea
Introducing the microscopic variable $\M\equiv \sqrt{N}M$, we find
\ba
	Z & \sim \int_{S^2} \rmd \overrightarrow{L_C}
	\int_{S^2} \rmd \overrightarrow{L_F}
	\int_{S^2} \rmd \overrightarrow{R_C}
	\int_{S^2} \rmd \overrightarrow{R_F}
	~\exp\bigg(
	- \frac{1}{2}\Tr[(\overrightarrow{R_C}\cdot \overrightarrow{\sigma})
	(\overrightarrow{L_C}\cdot \overrightarrow{\sigma})]
	\Tr[(\overrightarrow{R_F}\cdot \overrightarrow{\sigma})\M 
	(\overrightarrow{L_F}\cdot \overrightarrow{\sigma})\M^\dagger]
	\bigg)
	\\
	& = \int_{S^2} \rmd \overrightarrow{L_C}
	\int_{S^2} \rmd \overrightarrow{L_F}
	\int_{S^2} \rmd \overrightarrow{R_C}
	\int_{S^2} \rmd \overrightarrow{R_F}
	~\exp\Big(
	- (\overrightarrow{R_C}\cdot\overrightarrow{L_C})
	\Tr[(\overrightarrow{R_F}\cdot \overrightarrow{\sigma})\M 
	(\overrightarrow{L_F}\cdot \overrightarrow{\sigma})\M^\dagger]
	\Big)
	\\
	& = \int_{-1}^{1}\rmd x \int_{S^2} \rmd \overrightarrow{L_F}
	\int_{S^2} \rmd \overrightarrow{R_F}
	~\exp\Big(
	x \Tr[(\overrightarrow{R_F}\cdot \overrightarrow{\sigma})\M 
	(\overrightarrow{L_F}\cdot \overrightarrow{\sigma})\M^\dagger]
	\Big)\,.
\ea
\end{widetext}
In the special case $\M=\hat{m}\1_2$, we have 
\ba
	\Tr[(\overrightarrow{R_F}\cdot \overrightarrow{\sigma})\M 
	(\overrightarrow{L_F}\cdot \overrightarrow{\sigma})\M^\dagger]=2|\hat{m}|^2 \overrightarrow{R_F}\cdot \overrightarrow{L_F}\,,
\ea
hence
\ba
	Z & = \frac{1}{4} \int_{-1}^{1}\rmd x \int_{-1}^{1}\rmd y \exp\mkakko{2|\hat{m}|^2xy} 
	= \frac{\text{Shi}(2|\hat{m}|^2)}{2|\hat{m}|^2}\,,
\ea
where $\text{Shi}(x)$ is the hyperbolic sine integral \cite{dlmf_shi} and we adopted the normalization such that $Z=1$ in the chiral limit.

\subsection{Effective theory of the adjoint CFL phase}

Let us discuss the quark-mass dependence of the partition function in the adjoint CFL phase, which appears for $0.357<|v|<1.072$. In this phase, $\Omega^{L,R}=\rho \1_3$ is the ground state, where the scale $\rho$ is dynamically determined by $v$. Color and flavor fluctuations can then be parametrized as
\ba
	\Omega^L = \rho O^L ~~\text{and}~~ \Omega^R = \rho O^R ~~\text{with}~~O^{L,R}\in\SO(3)\,.
\ea
For brevity, we define $4\times 4$ matrices
\ba
	\mathbf{O}_L & \equiv \rho O^L_{AB}\sigma^A\otimes \sigma^B + v \1_2 \otimes \1_2\,,
	\\
	\mathbf{O}_R & \equiv \rho O^R_{AB}\sigma^A\otimes \sigma^B + v \1_2 \otimes \1_2\,.
\ea
Recalling \eqref{eq:924799}, we have for the large-$N$ partition function
\ba
	Z & \sim \int_{\SO(3)}\hspace{-4mm}\rmd O^L \int_{\SO(3)}\hspace{-4mm}\rmd O^R~
	{\det}^N \begin{pmatrix}
		\1_2 \otimes M & \mathbf{O}_R
		\\
		\mathbf{O}_L & \1_2 \otimes M^\dagger
	\end{pmatrix}
	\\
	& \propto 
	\int_{\SO(3)}\hspace{-4mm}\rmd O^L \int_{\SO(3)}\hspace{-4mm}\rmd O^R
	\notag
	\\
	& \quad \times 
	{\det}^N \kkakko{
		\1_4 - (\1_2\otimes M)\mathbf{O}_L^{-1}(\1_2\otimes M^\dagger)\mathbf{O}_R^{-1}
	}
	\\
	& \simeq 
	\int_{\SO(3)}\hspace{-4mm}\rmd O^L \int_{\SO(3)}\hspace{-4mm}\rmd O^R
	\notag
	\\
	& \quad \times 
	\exp\mkakko{
		- \Tr[(\1_2\otimes \M)\mathbf{O}_L^{-1}(\1_2\otimes \M^\dagger)\mathbf{O}_R^{-1}]
	}
	\label{eq:eepp}
\ea
where $\M\equiv \sqrt{N}M$. The evaluation of the trace in the exponent is a bit tricky. Fist of all, note that at the second order in quark masses, there are only two combinations of $O^L,O^R,M$ and $M^\dagger$ that are invariant under $\SU(2)_L\times\SU(2)_R$. The first invariant is $\Tr(M^\dagger M)$. To find the second invariant, note that $M$ and $M^\dagger$ do not carry color indices. Thus we must look at the color singlet product $(O^R)^T O^L$, which transforms as $(\mathbf{3},\mathbf{3})$ under $\SU(2)_L\times\SU(2)_R$. Since $\Tr(\sigma^A M \sigma^B M^\dagger)$ also transforms as $(\mathbf{3},\mathbf{3})$, we arrive at the second invariant
\ba
	[(O^R)^T O^L]_{AB} \Tr(\sigma^A M \sigma^B M^\dagger) \,.
	\label{eq:OOinv}
\ea 
Therefore the exponent in \eqref{eq:eepp} must be a linear combination of $\Tr(M^\dagger M)$ and \eqref{eq:OOinv}. To fix the coefficients of these two terms, we substituted simple forms 
\ba
	O^L & = \begin{pmatrix}
	\cos \theta_L & - \sin\theta_L & 0 \\ \sin \theta_L & \cos\theta_L & 0 \\ 0&0&1
	\end{pmatrix},
	\\
	O^R & = \begin{pmatrix}
	1 & 0 & 0 \\ 0 &\cos\theta_R&-\sin\theta_R \\ 0 &\sin\theta_R&\cos\theta_R
	\end{pmatrix},
\ea
and evaluated the trace in \eqref{eq:eepp} using a symbolic computation software \cite{10.7717/peerj-cs.103}. This enabled us to derive the formula
\ba
	& \Tr[(\1_2\otimes \M)\mathbf{O}_L^{-1}(\1_2\otimes \M^\dagger)\mathbf{O}_R^{-1}]
	\notag
	\\
	=\; & \frac{2}{(v+\rho)^2(v-3\rho)^2}\Big\{ (v-2\rho)^2 \Tr(\M^\dagger \M)
	\notag
	\\
	& + \rho^2 [(O^R)^T O^L]_{AB} \Tr(\sigma^A \M \sigma^B \M^\dagger) \Big\}\,.
	\label{eq:063}
\ea
To test \eqref{eq:063}, we randomly sampled $O^L$ and $O^R$ from $\SO(3)$ and evaluated both sides of \eqref{eq:063} numerically. We found that they match up to 15 digits, so we are pretty confident that \eqref{eq:063} is correct. 

Plugging \eqref{eq:063} into \eqref{eq:eepp} yields 
\ba
	Z & \sim \int_{\SO(3)}\hspace{-4mm}\rmd \hat{O}~
	\exp \big[
	\xi_1 \Tr(\M^\dagger \M) + \xi_2 \hat{O}_{AB} \Tr(\sigma^A \M \sigma^B \M^\dagger)
	\big]
	\label{eq:Z293794}
\ea
with 
\ba
	\xi_1 & \equiv - \frac{2(v-2\rho)^2}{(v+\rho)^2(v-3\rho)^2}\,,
	\\
	\xi_2 & \equiv - \frac{2\rho^2}{(v+\rho)^2(v-3\rho)^2}\,.
\ea
The variable $\hat{O}\equiv(O^R)^T O^L$ represents the Nambu-Goldstone mode arising from the spontaneous chiral symmetry breaking $\SU(2)_L\times\SU(2)_R\to\SU(2)_V$. Compared with the chiral Lagrangian of the usual CFL phase \cite{Son:1999cm,Son:2000tu,Schafer:2001za}, it is notable that \eqref{eq:Z293794} has no term proportional to $\M^2$ and $\M^{\dagger 2}$. This is not surprising, considering that the $\U(1)_A$ symmetry is unbroken in the adjoint CFL phase. 

In the special case $\M=\hat{m}\1_2$, we have \footnote{The formula $\int_{\SO(3)} \rmd \hat{O}~\exp(x\Tr\hat{O})=\rme^x[I_0(2x)-I_1(2x)]$ follows from eq.~(7.33) in \cite{Balantekin:2001id}.}
\ba
	Z & \sim \rme^{2\xi_1|\hat{m}|^2}
	\int_{\SO(3)}\hspace{-4mm}\rmd \hat{O}~
	\rme^{2\xi_2 |\hat{m}|^2 \Tr \hat{O}}
	\\
	& = \rme^{2(\xi_1+\xi_2)|\hat{m}|^2}\kkakko{I_0(4\xi_2|\hat{m}|^2)-I_1(4\xi_2|\hat{m}|^2)}
\ea
where $I_0$ and $I_1$ are the modified Bessel function of the first kind.

As a side remark, we note that if the mapping $\hat{O}_{AB}=\frac{1}{2}\Tr(U\sigma^A U^\dagger \sigma^B)$ between $U\in\SU(2)$ and $\hat{O}\in\SO(3)$ is used, then 
\ba
	\hat{O}_{AB} \Tr(\sigma^A \M \sigma^B \M^\dagger)
	& = 2 |\Tr(\M U)|^2 - \Tr(\M^\dagger \M)\,.
\ea
This type of mass term also arises in a chiral RMT discussed in \cite{Kanazawa:2016nlh}. However the qualitative difference between the present work and \cite{Kanazawa:2016nlh} must be underlined. The matrix model in \cite{Kanazawa:2016nlh} is heavy tailed, and has no explicit color structure in the Dirac operator. By contrast, the random matrices in this paper obey Gaussian distributions, and there is an explicit color structure in the Dirac operator \eqref{eq:defDD}.

\section{\label{sc:conc}Conclusions}

In this paper we investigated the formation of quark-antiquark condensates that dynamically break color and flavor symmetries. We constructed a new chiral random matrix model and showed, by taking the large-$N$ limit, that a novel \emph{adjoint CFL phase} appears for a particular range of the parameter of the model. Outside this range there appear a polar phase and a normal phase. We analytically evaluated the large-$N$ partition function and derived low-energy effective theories for soft modes in the polar phase and the adjoint CFL phase. We hope this work provides a deeper insight into both dense QCD and random matrix theory. 

For a technical reason we had to limit ourselves to two colors and two flavors in the second half of this paper. It would be quite worthwhile to explore the phase structure of this matrix model for the most interesting case of three colors and three flavors. This is left for future work.

\bibliography{draft_adjoint.bbl}
\end{document}